\documentclass[aps,prd,twocolumn,showpacs,superscriptaddress,groupedaddress]{revtex4-2}  
\usepackage[T1]{fontenc}
\usepackage{lmodern} 
\usepackage{graphicx}  
\usepackage{dcolumn}   
\usepackage{bm}        
\usepackage{amssymb}   
\usepackage{amsmath}
\usepackage{bbold}
\usepackage{array}
\usepackage{makecell}
\usepackage{braket}
\usepackage{longtable}
\usepackage{supertabular,booktabs}

\usepackage{titlesec} 
\usepackage[colorlinks,citecolor=blue,urlcolor=blue,hypertexnames=true]{hyperref}
\usepackage{subfigure}

\usepackage[usenames,dvipsnames,svgnames,table]{xcolor} 

\newcommand{\be}{\begin{equation}}
\newcommand{\ee}{\end{equation}}
\newcommand{\bea}{\begin{eqnarray}}
\newcommand{\eea}{\end{eqnarray}}
\newcommand{\bml}{\begin{subequations}}
\newcommand{\eml}{\end{subequations}}
\newcommand{\bfig}{\begin{figure}}
\newcommand{\efig}{\end{figure}}

\newcommand{\bmat}{\begin{pmatrix}}
\newcommand{\emat}{\end{pmatrix}}

\begin{document}

\widetext


\title{\textcolor{Sepia}{\textbf \huge\Large Circuit Complexity in an interacting quenched Quantum Field Theory}}

\author{{\large  Sayantan Choudhury${}^{1}$}}
\email{sayantan_ccsp@sgtu v+
+niversity.org,\\  sayanphysicsisi@gmail.com}

\author{ \large Rakshit Mandish Gharat${}^{2}$}
\email{rakshitmandishgharat.196ph018@nitk.edu.in }
\author{\large Saptarshi Mandal${}^{3}$}
\email{saptarshijhikra@gmail.com }
\author{ \large Nilesh Pandey${}^{4}$}
\email{nilesh911999@gmail.com  }

\affiliation{ ${}^{1}$Centre For Cosmology and Science Popularization (CCSP),\\
        SGT University, Gurugram, Delhi- NCR, Haryana- 122505, India,}
\affiliation{${}^{2}$Department of Physics, National Institute of Technology Karnataka, Surathkal, Karnataka-575025, India,}
\affiliation{${}^{3}$ Department of Physics, Indian Institute of Technology Kharagpur, Kharagpur-721302, India,}
\affiliation{${}^{4}$Department of Applied Physics, Delhi Technological University, Delhi-110042, India.}

\begin{abstract}
{In this work, we explore the effects of a quantum quench on the circuit complexity for a quenched quantum field theory having weakly coupled quartic interaction. We use the invariant operator method, under a perturbative framework, for computing the ground state of this system}. We give the analytical expressions for specific reference and target states using the ground state of the system. Using a particular cost functional, we show the analytical computation of circuit complexity for the quenched and interacting field theory. Further, we give a numerical estimate of circuit complexity with respect to the quench rate, $\delta t$ for two coupled oscillators. The parametric variation of the unambiguous contribution of the circuit complexity for an arbitrary number of oscillators has been studied with respect to the dimensionless parameter $(t/\delta t$). We comment on the variation of circuit complexity for different values of coupling strength, different number of oscillators, and even in different dimensions. 
\end{abstract}

\pacs{}
\maketitle
\section{\textcolor{Sepia}{\textbf{ \large  Introduction}}}
\label{sec:introduction}
{The quest to understand the fundamental laws of nature has driven research in both high energy physics and quantum information theory, leading to a remarkable interplay between these fields. By applying information-theoretic tools to the study of various quantum systems, researchers have made groundbreaking discoveries, revealing the deep connections between seemingly disparate areas of physics\cite{intro:merge1,intro:merge3,intro:merge4,intro:merge5,intro:merge6,intro:merge7,intro:merge8}. As we continue to explore the interconnections between these fields, we can look forward to new insights and discoveries that have the potential to transform our understanding of the universe.}

{Circuit complexity is a fundamental concept in quantum information theory that studies the computational resources required to solve a problem using quantum circuits. Formally, circuit complexity measures the minimum number of gates required to implement a given quantum computation, as a function of the size of the input. The study of circuit complexity plays a crucial role in designing efficient quantum algorithms and understanding the power and limitations of quantum computers.}

{Since Leonard Susskind and his collaborators proposed the use of circuit complexity to study the interior of black holes \cite{s1,s2,s3,s4,s5,s6,s7,s8}, this approach has been extended to the study of quantum field theories. Researchers have found that circuit complexity can provide a useful tool for characterizing the complexity of entangled states in these theories and understanding their dynamics \cite{j1,j2,j3,Sinha}.}


{In the realm of many-body physics, the study of quantum quenches has become increasingly important in recent years, as they offer a powerful way to drive systems out of equilibrium and explore their dynamics \cite{intro:ooe1,intro:ooe2,intro:ooe3}. In a quantum quench, a time-dependent parameter is suddenly or slowly varied, driving the system away from its ground state and potentially leading to thermalization. The study of entanglement in the context of quenched systems has been a key area of investigation \cite{intro:quench1,intro:quench2,intro:quench3,intro:quench4,intro:quench5,intro:quench6,intro:quench7,intro:quench8,Ghosh:2017nlk,Ghosh:2019yjh}.  Moreover, the measurement of circuit complexity in quenched systems has emerged as a valuable tool for quantifying the computational resources required to simulate their dynamics \cite{Heller,compq}. Together, these studies shed light on the fundamental principles of non-equilibrium dynamics in quantum systems, and may pave the way for the development of novel quantum technologies.}

{In dynamical systems, the time-dependent Schrödinger equation plays a crucial role in understanding their evolution over time. To tackle this problem, the Lewis-Resenfeld invariant-operator method has been developed, which allows one to determine the time-dependent eigenstates of such systems \cite{doi:10.1063/1.1664991}. Additionally, the method can be extended to consider the adiabatic evolution of time-dependent parameters \cite{intro:pert-adiab1,intro:pert-adiab2}, providing a means to compute time-independent perturbative corrections to the eigenstates \cite{2020FrP.....8..189C}. The exact form of time-dependent parameters in these eigenstates can be found by solving the Ermakov-Milne-Pinney equation, which can be efficiently computed using the Mathematica software. These methods offer a powerful means to investigate the behavior of dynamical systems, shedding new light on the intricate interplay between their time-dependent parameters and their evolution over time.}

{In this research article, we investigate the circuit complexity of an interacting (quartic) quenched quantum field theory using the invariant operator method as described in the appendix of our previous work \cite{QE1}. The quench protocol we employ is the most commonly studied in the literature, and we use the results of \cite{Sinha} based on Nielsen's geometric approach to compute the circuit complexity. Specifically, we focus on graphically representing the time evolution of the unambiguous contribution of circuit complexity under different parametric variations. Our findings contribute to the growing body of literature on circuit complexity in quantum field theory and provide insights into the dynamics of complex interacting systems. For more details on the chosen quench protocol and methodology, readers can refer to \cite{Caputa:2017ixa,Heller,Sinha}.} 


The organization of the paper is as follows:
\begin{itemize}
    \item Discretising, a Quantum Field Theory (QFT) with quartic interaction, on a lattice, we decouple the Hamiltonian using Fourier modes in section \ref{sec1}. Evidently, the decoupled Hamiltonian refers to that of $N$ coupled oscillators having a quartic perturbative coupling. The frequency of these oscillators is quenched by choosing a particular protocol.
    \item In section \ref{sec2}, we use the invariant operator method to compute the time-dependent ground states and also the first-order perturbative corrections to the ground state, of the quenched Hamiltonian. {Notably, our research article represents the first time that this method has been applied to the computation of circuit complexity in a quenched field theory. This innovative approach allows for a more comprehensive understanding of the dynamics of complex interacting systems and provides new insights into the behavior of circuit complexity under quenched conditions.}
    \item Using the ground, we fix a specific reference and target state in section \ref{sec3}. 
    {We then evaluate the circuit complexity of the chosen reference and target state in an interacting (quartic) quenched quantum field theory using a particular cost function. We also evaluate the continuum limit of circuit complexity. Our results are based on a modification of the results presented in \cite{Sinha}, which is founded on Nielsen's geometric approach \cite{n1,n2,n3,n4,n5}. The method we use is more general than the covariance matrix approach used in \cite{Heller} and is, therefore, applicable in a perturbative framework.} 
    \item In section \ref{sec4}, we numerically evaluate circuit complexity for different sets of parameters and comment on the dynamical behavior of the circuit complexity in three different regimes.
    \item Section \ref{sec5} encapsulates the conclusions we draw from the results obtained in this work. 
\end{itemize}

\section{\textcolor{Sepia}{\textbf{\large The Setup and the Quench protocol}}}\label{sec1} 
{While the effect of quench on free field theories has been previously studied, this is one of the first attempts to understand its impact in the context of interacting theory. To begin, we focus on a scalar field theory with $\hat{\lambda}\phi^4$ interaction term, which we regulate by placing it on a lattice. Once discretized, the Hamiltonian represents a family of $N$ coupled anharmonic oscillators. We transform the original coordinates to normal modes to decouple the Hamiltonian, which allows us to compute the eigenstates for the system in a simpler way. To facilitate comparison with previous works, we follow the notations used in \cite{Sinha,j1}. We also describe the time-dependent quench profile chosen, which is the frequency of these oscillators.}
The Hamiltonian for a scalar field theory with a $\hat{\lambda} \phi^4$ interaction is given by \cite{j1}, 
 \begin{align}
 \mathcal{H}=\frac{1}{2} \int d^{d-1} x&\Bigg[\pi(x)^{2}+(\nabla \phi(x))^{2}+m^{2} \phi(x)^{2}\nonumber\\
 &+\frac{\hat{\lambda}}{12}\phi(x)^{4}\Bigg]~. 
\end{align}
Here $d$ is the space-time dimensions. We assume that the coupling $\hat{\lambda}<<1$, so that we can work in a perturbative framework. This theory can be discretized on a $d-1$ dimensional lattice, {which is characterised by lattice spacing, $\delta$.} 

Closely following the prescription shown in \cite{j1, Sinha} by making proper substitutions one can show that the Hamiltonian for the scalar field theory having quartic interaction term can be expressed as,
\begin{equation}\label{infH}
\begin{split}
\mathcal{H}={}&\sum_{\vec{n}}\Big\{\frac{\hat{P}(\vec{n})^{2}}{2 M}+\frac{1}{2} M\Big[\omega^{2} \hat{X}(\vec{n})^{2}\\&+\eta^{2} \sum_{i}\left(\hat{X}(\vec{n})-\hat{X}\left(\vec{n}-\hat{x}_{i}\right)\right)^{2}\\
&+2\lambda \hat{X}(\vec{n})^{4}\Big]\Big\}.
\end{split}
\end{equation}
Here $\vec{n}$ denotes the spatial location of the points on the lattice,  $\hat{x}_{i}$ represents the unit vectors along the lattice while $\omega$ represents the frequency of individual oscillators and $\eta$ denotes inter-mass coupling.
The above Hamiltonian, in Eq.\eqref{infH} represents a family of infinite coupled anharmonic oscillators. 
We use normal mode coordinates as discrete Fourier transform of the original coordinates, given by:
\begin{equation} \label{Eq_4.2}
    x_a = \frac{1}{\sqrt{N}} \sum_{k=0}^{N-1} \exp{\left[i \frac{ 2\pi a }{N} k\right]} \tilde{x}_k 
\end{equation}
\begin{equation} \label{Eq_4.3}
    p_a = \frac{1}{\sqrt{N}} \sum_{k=0}^{N-1} \exp{\left[i \frac{ 2\pi a }{N} k\right]} \tilde{p}_k
\end{equation}
Setting $M=1$ for simplicity, the Hamiltonian of Eq.\eqref{infH} can be rewritten in normal modes as:
\bea\label{splitH}
H=H_k+H'_{\phi^4}\hspace{5pt},
\eea
where,
\bea
\label{decoup}
H_k=\frac{1}{2} \sum_{k=0}^{N-1} \Big [|\tilde{p}_k|^2 + \omega_k^2|\tilde{x}_k|^2\Big],
\eea
denotes the unperturbed (free) Hamiltonian which can be decoupled for each of the $N$ oscillators.
Here,
\begin{equation}\label{freq}
  \omega_k^2=\omega^2 + 4 \eta^2 \sin^2{\Big(\frac{\pi k}{N}\Big)},
\end{equation}
denotes the freqency for each of the $N$ oscillators.
The exact form of the eigenstates for the unperturbed Hamultonian of Eq.\eqref{decoup} has been computed in the subsection \ref{subsec3.1}.

On the other hand the $\lambda\phi^4$ perturbation term in the Hamiltonian of Eq.\eqref{splitH} can be dealt with by transforming the form of perturbations in normal modes:
\begin{equation}\label{pert}
   \begin{aligned}
   H'_{\phi^4} = \frac{\lambda} {N} \sum_{k_1,k_2,k_3 = 0}^{N-1} \tilde{x}_{\alpha} \tilde{x}_{k_1} \tilde{x}_{k_2} \tilde{x}_{k_3} \ ; \\\ \alpha = N-k_1-k_2-k_3\ \text{mod}\ N
   \end{aligned}
\end{equation}
The contribution of the above Hamiltonian in Eq.\eqref{pert} is evaluated by approximating the first order correction to the eigenstates of unperturbed Hamiltonian by employing the use of time-independent perturbation theory in the subsection \ref{perturbed}.\\
We now consider the frequency $\omega$ in, Eq.\eqref{freq} as a time-dependent quench profile.
One of the most common quench profiles used in literature \cite{Caputa:2017ixa} is given by:
\be\label{quenchprof}
\omega^2(t/\delta t)=\omega_0^2\left[\tanh^2{\left(\frac{t}{\delta t}\right)}\right].
\ee  
Here $\omega_0$ can be considered as a free parameter and $\delta t$ measures the quench rate. We choose this particular quench profile chosen since it admits an exact solution for the mode functions given in \cite{Caputa:2017ixa}. Note that this profile attains a constant value at very early and late time. Also, for this chosen form of quench profile, the dynamical changes in the system occur in the time window $[-\delta t,\delta t]$. We will set $t/\delta t=T$ and $\omega_0=1$. 
The respective frequencies in the normal mode basis take the following form,
\begin{equation}\label{frqe}
  \omega_k=\sqrt{\omega(T)^2 + 4 \eta^2 \sin^2{\Big(\frac{\pi k}{N}\Big)}},  
\end{equation}
where $\omega(T)$ is the quench profile in Eq.\eqref{quenchprof} and $k$ runs from $0$ to $N-1$.\\
As the frequency of each oscillator now depends on time, the unperturbed Hamiltonian is evidently time-dependent. We employ the use of invariant operator method to compute the exact form of the unperturbed Hamiltonian. We emphasise that the perturbed Hamiltonian is not time-dependent and hence can be used as a time-independent perturbation applied to $N$ coupled oscillators. Using the ground state of total Hamiltonian of Eq.\eqref{splitH} we construct the reference as well as target states which are further used to evaluate the circuit complexity of this interacting quench model. 
\section{\textcolor{Sepia}{\textbf{\large Constructing Wave function for a $\phi^4$ quench model}}} 
\label{sec2}
In this section our prime objective is to derive an analytical expression for the eigenstates of Hamiltonian in Eq.\eqref{splitH}, by using Lewis-Resenfield invariant opertor method and approximate it to the first order perturbative correction. The expression for eigenstates of decoupled and unperturbed Hamiltonian in Eq.\eqref{decoup} is derived using invariant operator method, in the subsection \ref{subsec3.1}. The first order perurbative correction to the ground state of the decoupled Hamiltonian, is derived in the subsection \ref{perturbed}.

\subsection{Eigenstates and Eigenvalues for unperturbed Hamiltonian}\label{subsec3.1}
As shown earlier, in the normal mode basis, the unperturbed Hamiltonian, of Eq.\eqref{decoup},  for $N$ oscillators decouples. The wavefunction for any generic state of the $N$ oscillators is then a product of eigenstates for each decoupled Hamiltonian, of Eq.\eqref{decoup}:
\bea
\psi_{n_0\cdots n_{N-1}}\left(\tilde{x}_{0}, \cdots \tilde{x}_{N-1},T\right)=\psi_{n_0}(\tilde{x}_0,T)\psi_{n_1}(\tilde{x}_1,T)\nonumber\\\cdots\psi_{n_{N-1}}(\tilde{x}_{N-1},T).
\eea
Here $T$ denotes the time-dependence of eigenstates emerging due to the quenched frequency from Eq.\eqref{quenchprof}. We deal with these time-dependent eigenstates by employing the use of invariant operator method, closely following the prescription of \cite{1994PhRvA..50.1035Y}. In Appendix \ref{app:Invar}, we have briefly mentioned the steps one can follow to compute an analytical expression for eigenstates of a quenched Hamiltonian of Eq.\eqref{decoup} using invariant operator method. 
Using Eq.\eqref{eigenI} and the arguments given in Appendix \ref{app:Invar} one can show that the expression for the eigenstate of total unperturbed Hamiltonian of $N$ coupled oscillators is:
\begin{widetext}
    \bea\label{nth}
     \psi^{(0)}_{n_1\cdots n_{N-1}}=\left(\frac{1}{2^{n_0+n_1+\cdots+n_{N-1}}n_0!\cdots n_{N-1}!}\right)\left(\frac{g_0g_1\cdots g_{N-1}}{\pi^N}\right)^{1/4}\exp\left[-\frac{i}{2}\sum_{k=0}^{N-1}(2n_k+1)\gamma_k\right]\exp\left[-\frac{1}{2}\sum_{k=0}^{N-1}\tilde{\nu}_k\tilde{x}_k^2\right]\nonumber\\
     \times\textbf{H}_{n_0}\left[\sqrt{\dot{\gamma_0}}\tilde{x}_0\right]\cdots\textbf{H}_{n_{N-1}}\left[\sqrt{\dot{\gamma}_{N-1}}\tilde{x}_{N-1}\right].~~~~~
     \eea
\end{widetext}
Here $g_k=\dot{\gamma}_k$ and,
\begin{equation}
    \tilde{\nu}_k=\dot{\gamma}_k\left(1-\frac{i\dot{\rho_k}}{\rho_k\dot{\gamma_k}}\right),
\end{equation}
for $k=0,1\cdots,N-1$. All other symbols used in Eq.\eqref{nth} are defined in Appendix \ref{app:Invar}. In this work, we focus on the ground state of the wavefunction shown in Eq.\eqref{nth}, which can be written as:
    \bea\label{gnd}
     \psi^{(0)}_{0\cdots0}=\left(\frac{g_0g_1\cdots g_{N-1}}{\pi^N}\right)^{1/4}\exp\left[-\frac{1}{2}\sum_{k=0}^{N-1}(i\gamma_k +  \tilde{\nu}_k\tilde{x}_k^2)\right].\nonumber\\
     \eea
To compute the eigenvalues of the unperturbed quenched Hamiltonian in Eq.\eqref{decoup}
one can again use the invariant operator method. From the arguments given in  \cite{2020FrP.....8..189C} the energy eigenvalues for time-dependent harmonic oscillators can be evaluated by multiplying a time-dependent factor by the expression of energy eigenvalues of time-independent harmonic oscillators.  Hence, one can show that the energy eigenvalues for each of the $N$ decoupled oscillators become:
\begin{equation}
 \begin{aligned}
    \bra{\psi_{n_k}}H_i\ket{\psi_{n_k}}&=W_k(T)\left[n_k+\frac{1}{2}\right].
 \end{aligned}
\end{equation}
Here $W_k(T)$ for $k=0,\cdots,N-1$ is a time-dependent factor for each oscillator given by,
\be\label{eigent}
W_k(T)=\frac{\dot{\gamma_k}}{2}\left(\frac{\dot{\rho_k}+\rho_i^2\rho_i^2+\rho_i^2\dot{\gamma_i}}{\rho_i^2\dot{\gamma_i}^2}\right).
\ee
Using the above form of time-dependent eigenvalues, one can compute the energy eigenvalues for the decoupled Hamiltonian of Eq.\eqref{decoup}, which is given by:
\bea
\bra{\psi_{n_1,\cdots,n_{N-1}}^{(0)}}H\ket{\psi_{n_1,\cdots,n_{N-1}}^{(0)}}=\sum_{k=0}^{N-1}W_k\left(n_k+\frac{1}{2}\right).\nonumber\\
\label{evalue}
\eea
The above expression for the eigenvalues of unperturbed Hamiltonian can now be used to approximate the first order time-independent perturbative correction to the ground state of Eq.\eqref{gnd}.

\subsection{Wavefunction for $\lambda\phi^4$ perturbation applied to the ground state of $N$ quenched-coupled oscillators}\label{perturbed}
In this section our prime objective is to evaluate analytical expression for the wavefunction of ground state of $N$ coupled oscillators in a perturbative framework, approximated to first order. We consider $\psi^{(1)}$ to be the first order correction arising due to Hamiltonian in Eq.\eqref{pert}. Hence, using Eq.\eqref{gnd}, the expression for ground state of total Hamiltonian Eq.\eqref{splitH} corrected to first order in $\lambda$ can be written as,
\begin{equation}\label{T1}
\begin{aligned}
   \psi_{0,0, \cdots 0}\left(\bar{x}_{0}, \cdots \tilde{x}_{N-1}\right)=\left(\frac{g_0g_1\cdots g_{N-1}}{\pi^N}\right)^{1/4}~~~~~~~\\\times\exp\left[-\frac{1}{2}\sum_{k=0}^{N-1}(i\gamma_k +  \tilde{\nu}_k\tilde{x}_k^2+\lambda\psi^{1})\right].
   \end{aligned}
\end{equation}
We take note of the fact that for $N$ coupled oscillators the $\lambda\phi^4$ perturbation can give rise to a combination of five terms viz., $x_a^4, x_b^2x_c^2, x_dx_e^3, x_fx_g^2x_h$ and $x_ix_jx_kx_l$. Hence, we have expressed the form of first order correction by closely following the notations used in \cite{Sinha},
\begin{widetext}
\begin{equation}\label{PC1}
\begin{aligned}
\psi_{4}^{1} =
&~~~\sum_{\substack{a=0\\ 4a \bmod N\equiv0}}^{N-1} B_{1}(a) ~~+ \sum_{\substack{b,c=0\\(2 b+2 c) \bmod N \equiv 0\\b \neq c} }^{N-1} \frac{B_{2}(b, c)}{2}+\sum_{\substack{d,e=0\\(3 e+d) \bmod N\equiv 0 \\d \neq e } }^{N-1} B_{3}(d, e)&~~~~~~~~~~~~~~~~~~~~\\ 
+ &\sum_{\substack{f,m,h=0\\(f+2 m+h) \bmod N\equiv0\\  f \neq m \neq h} }^{N-1} \frac{B_{4}(f, m, h)}{2} ~~+ 
\sum_{\substack{i,j,k,l=0\\(i+j+k+l) \bmod N\equiv 0\\i \neq j \neq k \neq l} }^{N-1} \frac{B_{5}(i,j,k,l)}{24}.&\\
\end{aligned}
\end{equation}
\end{widetext}
One can compute the exact form of coefficients for each of the five different perturbative terms by first choosing appropriate number of oscillators and then generalising the result for $N$ oscillators. Further one can compute the perturbative correction by setting $V$ as each perturbative term mentioned above, and then using the formula given below,
\begin{widetext}
\be\label{PC2}
\psi_{0,\cdots0}^{(1)}=\sum_{(n_0\cdots n_{N-1})\neq(0,\cdots0)}\frac{\bra{\psi_{n_0,\cdots n_{N-1}}^{(0)}}V\ket{\psi_{0,\cdots0}^{(0)}}\times\psi_{n_0,\cdots n_{N-1}}^{(0)}}{\bra{\psi_{0,\cdots0}^{(0)}}H\ket{\psi_{0,\cdots0}^{(0)}}-\bra{\psi_{n_0,\cdots n_{N-1}}^{(0)}}H\ket{\psi_{n_0,\cdots n_{N-1}}^{(0)}}}~~.
\ee
\end{widetext}
For example, if one wants to get the form of $B_3(d,e)$ in Eq.\eqref{PC1}, set number of oscillators to $N=2$ and put $V=x_0x_1^3$ in Eq.\eqref{PC2}. The form of the perturbative expansion thus obtained can be generalised for arbitrary number of, $N$ oscillators. We then repeat these steps to fix all the coefficients of perturbative expansion. The exact form of all these coefficients, using Eq.\eqref{PC2} is tabulated in Appendix \ref{app:Table}.

\section{\textcolor{Sepia}{\textbf{\large Analytical calculation for Circuit Complexity of $\phi^4$ quench model}}}\label{sec3}
The ground state of the total Hamiltonian, calculated in the previous section, given by Eq.\eqref{T1} is used to construct the reference and target states in the subsection \ref{sub:R/T}. Choosing a specific cost functional, we have derived the analytical expression for circuit complexity, by modifying the results of \cite{Sinha}, in the subsection \ref{sub:comp}. 

\subsection{Constructing Target/Reference states}\label{sub:R/T}
In the wavefunction for $N$ oscillators with quartic perturbation shown in Eq.\eqref{T1} following the prescription given in \cite{Sinha} one can write the exponent in the form of a matrix conjugated by a basis vector, $\vec{v}$. The wavefunction then takes the below given form:
\begin{equation}\label{compact}
\psi_{0,0, \cdots 0}^{s}\left(\tilde{x}_{0}, \cdots ,\tilde{x}_{N-1}\right) \approx\mathcal{N}^{s} \exp \left[-\frac{1}{2} v_{a}  A_{a b}^{s}  v_{b}\right].
\end{equation}
Here $\mathcal{N}^{s}$ denotes the normalisation factor and $A^s$ denotes a block diagonal matrix, for the respective state. Further, the space of circuits is parameterised by setting value of the running parameter $s$. At $s=1$ the above form of wavefunction coincides with the wavefunction in Eq.\eqref{T1} such that $\mathcal{N}_{s=1}$ becomes the normalising factor of Eq.\eqref{T1}, by an appropriate choice of the basis $\vec{v}$. Then $\psi_{0,0, \cdots 0}^{s=1}$ is referred to as the target state. There are many possible choices for choosing bases so as to obtain the terms in perturbative expansion in Eq.\eqref{PC1}. However as a minimal choice we choose the below mentioned basis,
\be\label{basis}
\vec{v} = \{\tilde{x}_{0}, \cdots \tilde{x}_{N-1},\tilde{x}_{0}^2,\cdots ,\tilde{x}_{N-1}^2, \cdots ,\tilde{x}_{a}\tilde{x}_{b}, \cdots\}.
\ee
In this basis one can show that the matrix $A$, in Eq.\eqref{compact} has a block diagonal form:
\begin{equation}
   A_{ab}^{s=1} = \begin{pmatrix}
    A_1 & 0 \\
    0 &  A_2
    \end{pmatrix}.
\end{equation}
$A_1$ contains coefficients of terms like $x_a^2$ and $x_a x_b$ in Eq.\eqref{T1} multiplied by $-2$. All the elements of $A_1$  can be fixed to obtain a specific form of target state. This block is often referred to as the \textit{unambiguous} block. 

The elements of $A_2$ block consist of coefficients which are basis dependent i.e. there is not any unique choice of basis vector for defining the elements of $A_2$ block because unlike $A_1$ block which only consist of coefficients of quadratic terms and they can be defined uniquely without any ambiguity, the $A_2$ block consists of elements which are coefficients of terms like ${\tilde{x}^2_{a}\tilde{x}^2_{b}}$, $\tilde{x}^2_{a}\tilde{x}^2_{b}\tilde{x}^2_{c}$, ${\tilde{x}_{a}\tilde{x}_{b}\tilde{x}_{c}\tilde{x}_{d}}$ which can be defined in several ways.
Due to this arbitrariness, the complexity for the ambiguous block will be different for different choices of basis. One cannot therefore fix elements of $A_2$ such that the contribution of $A_2$ to total complexity of the system is independent of choice of basis. Due to these ambiguities, the $A_2$ block is often referred to as \textit{ambiguous} block.
We construct the reference state by choosing the value of all the frequencies for each of the $N$ oscillators as $\tilde{\omega}_{ref}$. Since all the oscillators have same frequency, the reference state evidently would be time-independent. Then wavefunction in Eq.\eqref{PC1} can be modified to that of reference state, mentioned below,
\begin{equation}
    \psi^{s=0}(x_1,x_2,....,x_n) = \mathcal{N}^{s=0} \exp{\Big[-\sum_{i=0}^{N-1}\frac{\tilde{\omega}_{ref}}{2}\big(x_{i}^2+\lambda^0 x_{i}^4)\Big]}.
\end{equation}
Note that, $\lambda^0$ is a parameter denoting the non-Gaussian nature of the reference state and is not to be confused with the perturbative coupling $\lambda$ used in target state. In normal modes the above expression can be recast as shown below,
\begin{equation}\label{RS}
    \psi^{s=0}(\tilde{x}_1,\tilde{x}_2,....,\tilde{x}_n) = \mathcal{N}^{s=0} \exp{\Big[-\frac{1}{2}\Big(v_{a} A_{ab}^{s=0}v_b\Big)\Big]},
\end{equation} 
where the matrix $A_{ab}^{s=0}$ can be fixed as:
\begin{equation}
     A_{ab}^{s=0} = \begin{pmatrix}
    \tilde{\omega}_{ref}\mathbb{I}_{N\times N} & 0 \\
    0 &  A_2^{s=0}
    \end{pmatrix}.
\end{equation}
Again there will be ambiguities in fixing the elements of $A_2^{s=0}$ for the reasons already mentioned above.\\
Equipped with the target and and reference states for the quenched and interacting oscillators we can proceed to give an analytical expression for circuit complexity by getting around the ambiguities in the upcoming subsection.

\subsection{Analytical calculation of the complexity functional}\label{sub:comp}
In this subsection we outline analytical steps to compute the expression for circuit complexity for the previously mentioned target state Eq.\eqref{T1} starting with a reference state Eq.\eqref{RS} using the results of \cite{Sinha}.\\
As shown in \cite{Heller} the complexity functional depends on the chosen cost function. In this article we work with the following cost function,
\begin{equation} \label{4.16}
    \mathcal{F}_{\kappa}(s) = \sum_I p_I |Y^I|^{\kappa}.
\end{equation}
As shown in \cite{Sinha,cc1} the circuit complexity with this particular cost function becomes,
\begin{equation}
    \mathcal{C}_{\kappa} = \int_{s=0}^{1} \mathcal{F}_{\kappa} \ ds.
\end{equation}
Next, we write the complexity as sum of two terms,
\be
 \mathcal{C}_\kappa = \mathcal{C}_{\kappa}^{(1)} + \mathcal{C}_{\kappa}^{(2)}.
\ee
Here $\mathcal{C}_{\kappa}^{(1)}$ refers to the contribution to the circuit complexity from $A_1$ block while $\mathcal{C}_{\kappa}^{(2)}$ refers to the contribution from $A_2$ block. $\mathcal{C}_{\kappa}^{(1)}$ and $\mathcal{C}_{\kappa}^{(2)}$ can be given as ratio of eigenvalues of the respective blocks for chosen target \eqref{T1} and reference states Eq.\eqref{RS} as prescribed in \cite{Sinha,Heller},
\be\label{CE}
 \mathcal{C}_\kappa=\frac{1}{2^\kappa} \sum_{i=0}^{N-1}\Big|log\Big(\frac{\Lambda_i^{(1)}}{\tilde{\omega}_{ref}}\Big) \Big|^{\kappa}+\mathcal{A}\sum_{j}\Big|log\Big(\frac{\Lambda_j^{(2)}}{h_i\tilde{\omega}_{ref}\lambda_0}\Big) \Big|^{\kappa}.
\ee
Here $\mathcal{A}$ denotes the penalty factor.
Due to the ambiguities arising while fixing the form of $A_2$ block, numerically one cannot fix the form of $\Lambda_i^{(2)}$. However, as discussed in the Appendix \ref{app:Twoosc} we can choose a minimal basis such that elements of $A_2$ are fixed for $N=2$ oscillators. In the Appendix \ref{app:Twoosc}, we have computed the total circuit complexity for $N=2$ coupled oscillators having a quenched Hamiltonian with a quartic coupling.\\

In this work, have neglected the contribution of the ambiguous block i.e. $\mathcal{C}_{\kappa}^{(2)}$ as we cannot find any basis to get the numerically exact contribution of $A_2$ block for arbitrary number of oscillators. Nonetheless, we attempt to give an analytical form of $\mathcal{C}_{\kappa=1}^{(2)}$ in terms of renormalized parameters in the Appendix \ref{app:Renorm} again following the steps shown in \cite{Sinha}.\\

\subsection{The Continuum Limit for $\mathcal{C}_1$}\label{conti}
Now, we will compute the exact form of eigenvalues of $A_1$ block, taking the continuum limit. To find the form of $\Lambda_i^{(1)}$, we now reinstate the factor of $M$ previously set to $M=1$ in the section \ref{sec1}. In light of this the Hamiltonian, given in Eq.\eqref{splitH} will change, although retaining the previous form of eigenvalues and eigenfunctions. The new Hamiltonian with a factor of $M$ becomes,
\begin{equation}
\begin{aligned}
H= \frac{1}{M}\sum_{\vec{n}}\Big\{\frac{P(\vec{n})^{2}}{2 }+\frac{1}{2} M^2\Big[\omega^{2} X(\vec{n})^{2}+\Omega^{2} \sum_{i}(X(\vec{n})\\-X(\vec{n}-\hat{x}_{i}))^{2}+2\big\{\lambda_4 X(\vec{n})^{4}\}\Big]\Big\}.
\label{Eq_3.3}
\end{aligned}
\end{equation}
Considering the reinstated factor of $M$, we rescale some of the parameters as shown below,
\begin{align*}
\omega \rightarrow \frac{\omega}{\delta};\quad
\eta \rightarrow \frac{\eta}{\delta};\quad
\lambda \rightarrow \frac{\lambda}{\delta^2};\quad
\tilde{\omega}_{ref} \rightarrow \frac{\tilde{\omega}_{ref}}{\delta};\quad
\lambda^0 \rightarrow \frac{\lambda^0}{\delta}.
\end{align*}
Using these rescaled parameters one can generalise the form of eigenvalues $\Lambda_i^{(1)}$ using Mathematica by considering trial cases for different values of $N$ and eliminating all the factors aside for $\rho_k$ and $\tilde{\omega}_k$ by using appropriate formulae mentioned in previous sections. Below we show the generalised formula for eigenvalues of $A_1^{(s=1)}$ block depending on whether the chosen number of oscillators, $N$ is even or odd,
\begin{widetext}
    \begin{equation}  \label{5.2}
\begin{split}
   \Lambda _i=\frac{3 \lambda  \rho _i^2}{2N} \left(\frac{g_{\alpha }+2 g_i}{g_{\alpha} \left(\rho _i^2 \left(g_i+\omega _i^2\right)+\dot{\rho _i}\right)}-\frac{2 g_i \rho _{\alpha }^2}{\rho _i^2 \left(\rho _{\alpha }^2 \left(\omega _i^2 g_{\alpha }+g_i \left(2 g_{\alpha }+\omega _{\alpha }^2\right)\right)+g_i \dot{\rho _{\alpha }}\right)+\dot{\rho _i} g_{\alpha } \rho _{\alpha }^2}\right)+\frac{ \tilde{\nu} _i}{2 }, \quad N \text{: Even} \\
   =\frac{3 \lambda  g_i^2 \rho _i^4 \left(\left(g_i+\omega _i^2\right) \rho _{i}^2+\dot{\rho _i}\right)}{N g_i \left(\rho _i^2 \left(g_i+\omega _i^2\right)+\dot{\rho _i}\right) \left(\rho _i^2 \left(\rho _i^2 \left(g_i \omega _i^2+g_i \left(2 g_i+\omega _i^2\right)\right)+g_i \dot{\rho _i}\right)+g_i \dot{\rho _i} \rho _i^2\right)}+\frac{\tilde{\nu} _i}{2}, \quad N \text{: Odd}
\end{split}
\end{equation}
\end{widetext}
where the index, $\alpha=|N/2-i|$.
One can insert this expression for eigenvalues in Eq.\eqref{CE} to get the desired value of circuit complexity. We emphasize that the form of these eigenvalues make the circuit complexity a time-dependent quantity due to the chosen quench profile. This time-dependence of complexity is explored in numerical plots by choosing an appropriate scale of time.\\
One can check the behaviour of circuit complexity, $C_1$ at the continuum limit: $N\rightarrow\infty$ while $\delta\rightarrow0$ such that $L=(N\delta)$ is finite.\\
Now, in arbitrary dimensions, the equation for $\mathcal{C}^{(1)}_{\kappa=1}$, can be rewritten as: 
\be\label{CEd}
 \mathcal{C}^{(1)}_{\kappa=1}=\frac{1}{2} \sum_{k=1}^{d-1}\sum_{i=0}^{N-1}\Big|log\Big(\frac{\Lambda_i^{(1)}}{\tilde{\omega}_{ref}}\Big)\Big|.
\ee
For simplicity, if one chooses to set same frequency (of the respective oscillator) in each dimension, $d$, then one can write:
\begin{equation}\label{DIm}
    \omega_i=\sqrt{\frac{1}{{d-1}}\left[4(d-1)\eta^2\sin ^2\left(\frac{\pi i}{N}\right)+\omega_0^2 \tanh ^2\left(\frac{t}{\delta t}\right)\right]}.
\end{equation}
Hence, in arbitrary dimensions $d$, the circuit complexity can be still be computed using Eq.\eqref{CE}, thus getting rid of the lattice sums in Eq.\eqref{CEd} such that frequencies are set to Eq.\eqref{DIm}.

\begin{figure*}[htb!]
	\centering

		\includegraphics[width=16cm,height=10cm]{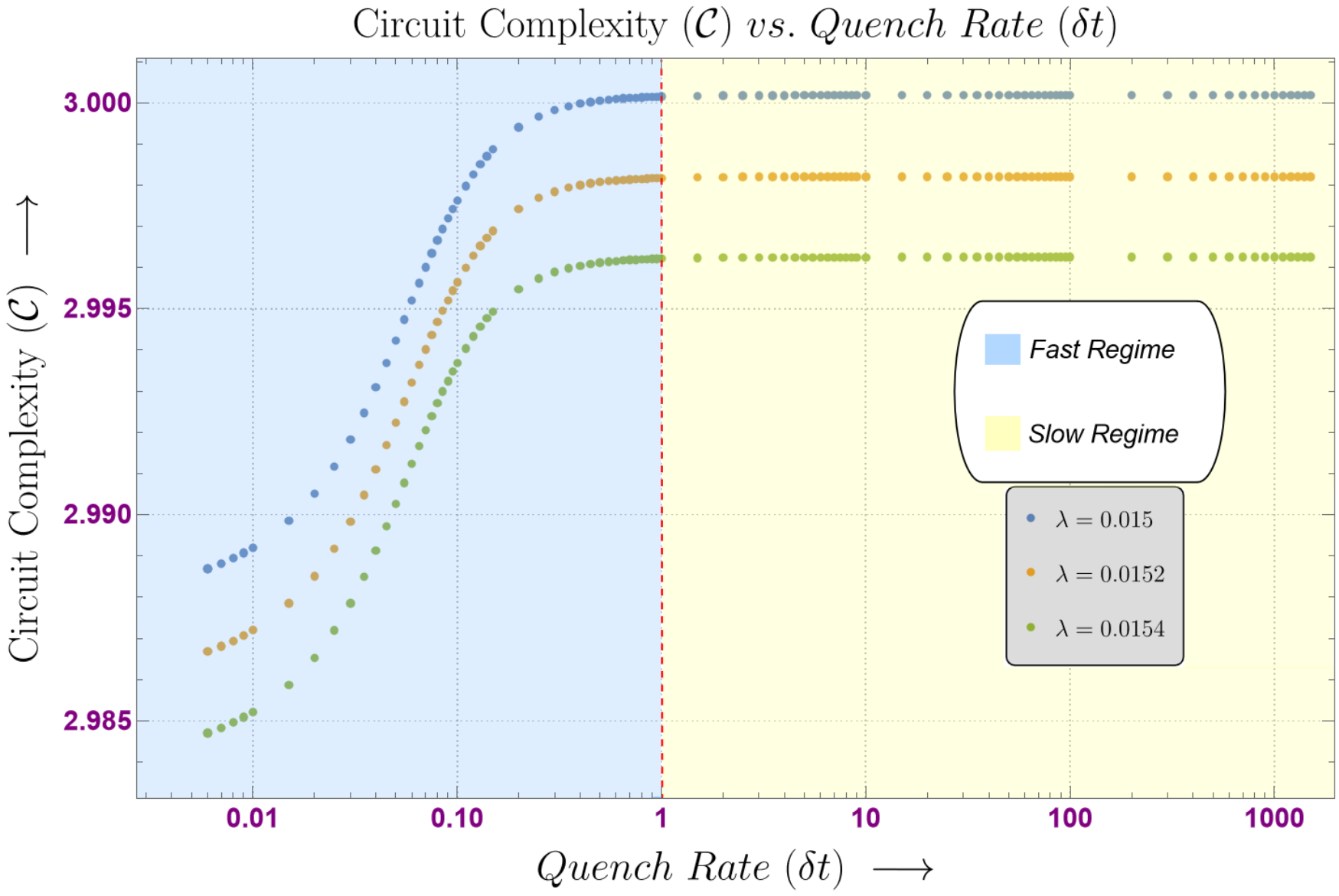}
	\caption{Log-Log variation of the total circuit complexity ($C$) with respect to the quench rate ($\delta t)$ for different orders of the coupling constant $\lambda$ for {two coupled oscillators} with quartic perturbation.}
	\label{fig:1}
\end{figure*}

\begin{figure*}[htb!]
	\centering

		\includegraphics[width=16cm,height=10cm]{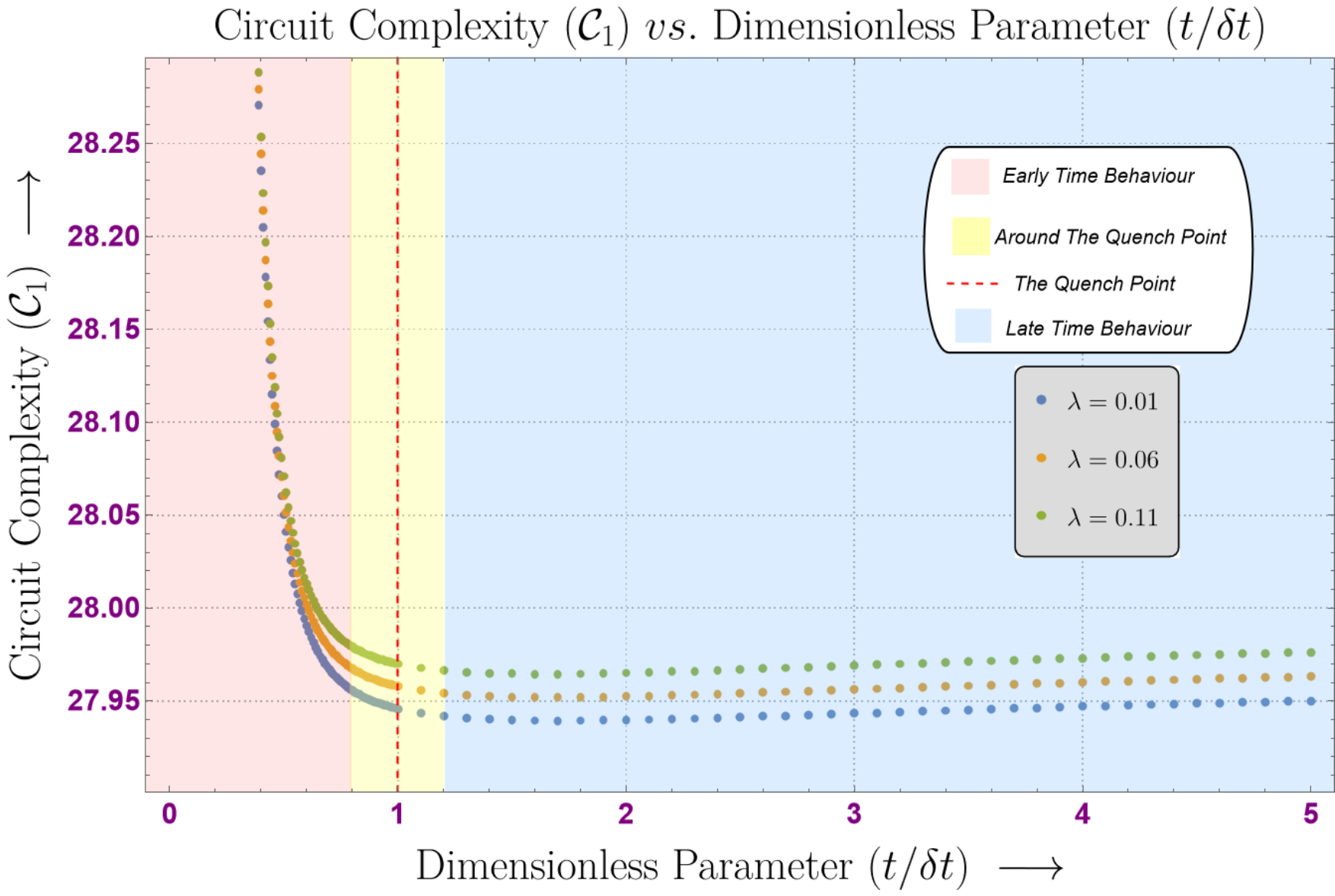}

	\caption{Variation of the circuit complexity for $A_1$ block ($C_1$) for $N=10,$ with respect to the dimensionless parameter ($t/\delta t)$ for different orders of the coupling constant $\lambda$.}
	\label{fig:2}
\end{figure*}
\begin{figure*}[htb!]
	\centering

		\includegraphics[width=16cm,height=10cm]{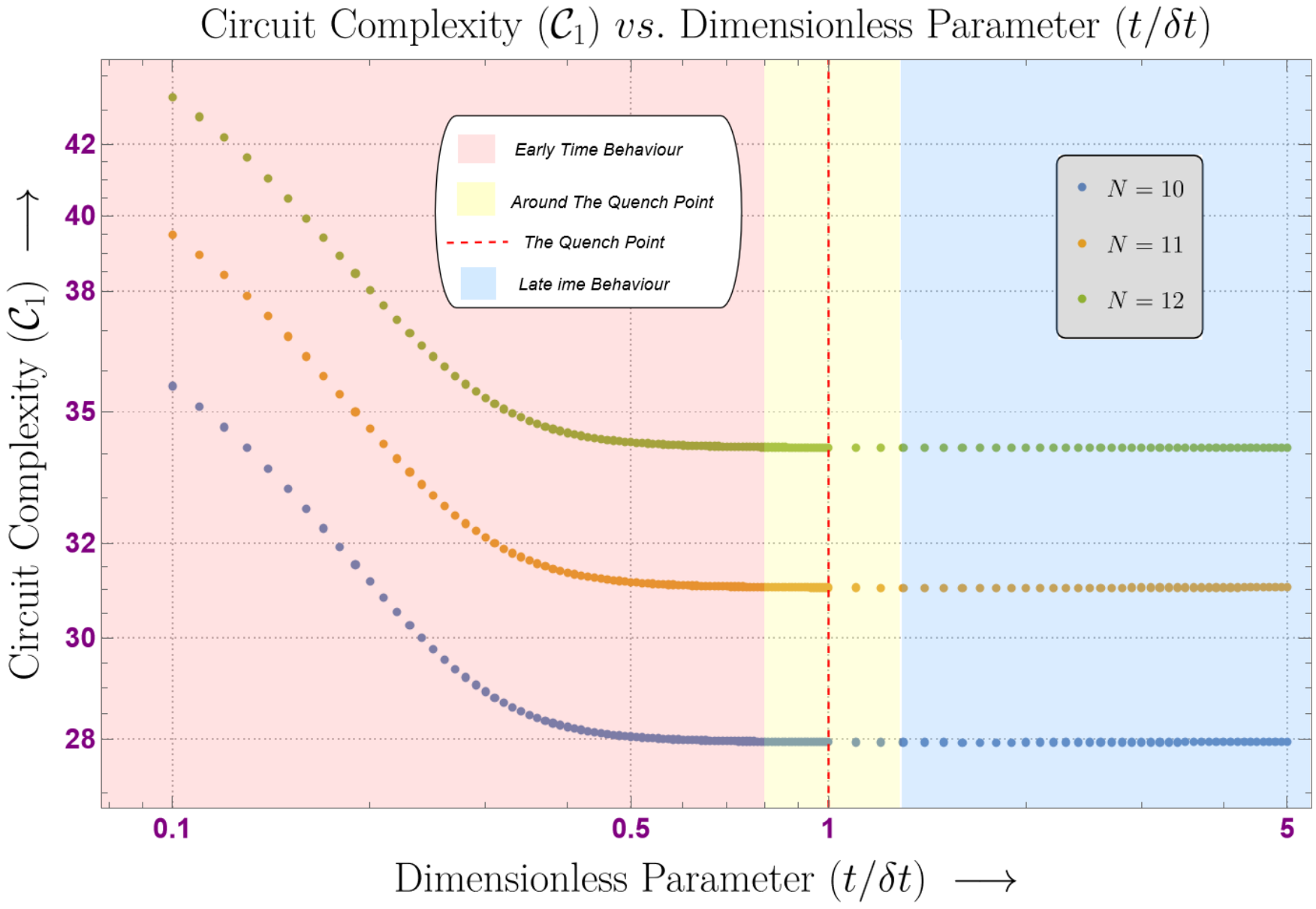}

	\caption{Semi-Log variation of the circuit complexity for $A_1$ block ($C_1$) at $\lambda=0.01,$ with respect to the dimensionless parameter ($t/\delta t)$ for different values of $N$.}
	\label{fig:3}
\end{figure*}

\begin{figure*}[htb!]
	\centering

		\includegraphics[width=16cm,height=10cm]{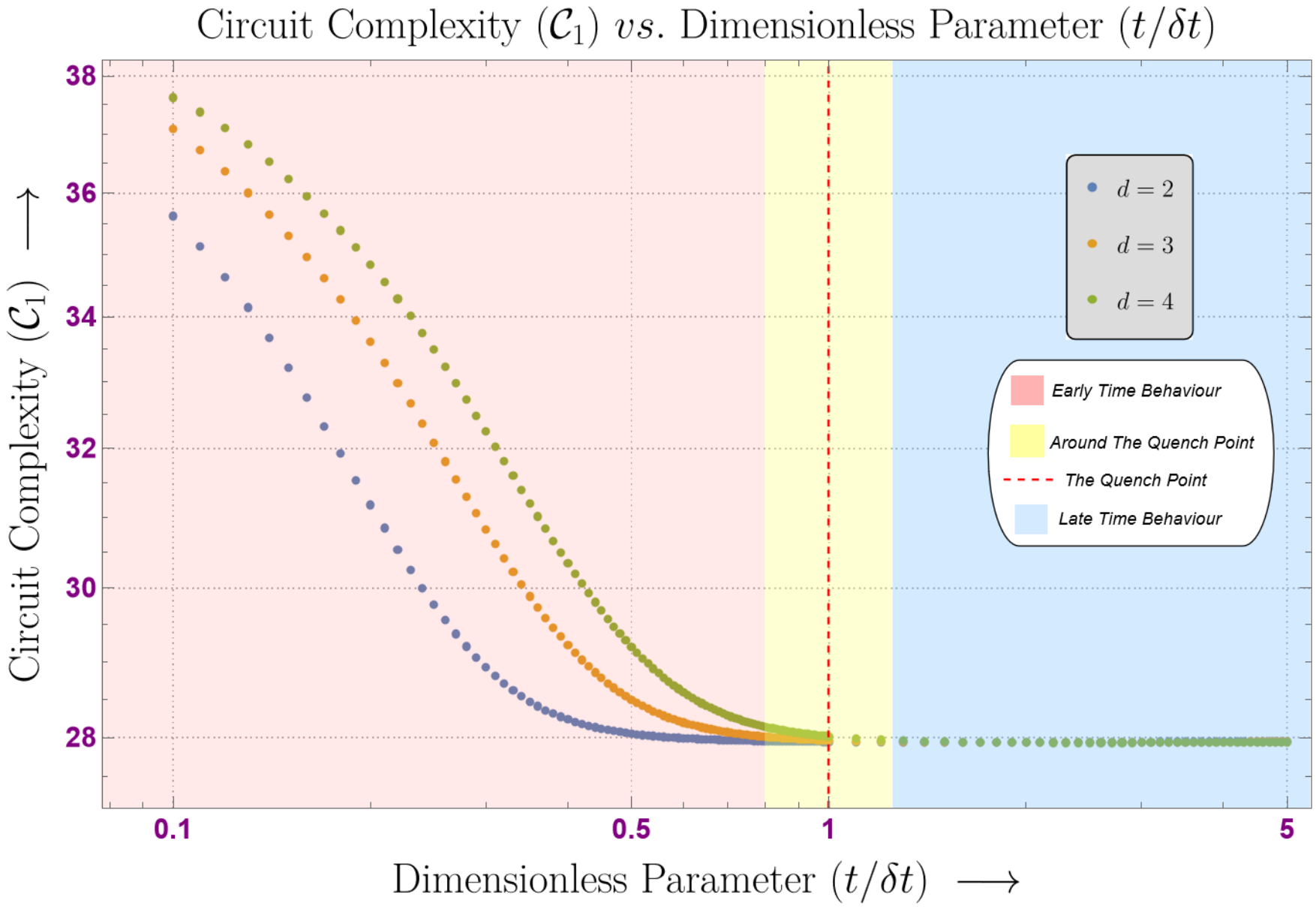}
		\caption{Semi-Log variation of the circuit complexity for $A_1$ block ($C_1$) at $\lambda=0.01, N=10$ with respect to the dimensionless parameter ($t/\delta t)$ for different dimensions $d$.}
		\label{fig:4}
\end{figure*}

\section{\textcolor{Sepia}{\textbf{ \large Numerical Results}}}\label{sec4}
In this section we numerically evaluate the value of circuit complexity for coupled oscillators using the expressions computed for total complexity of two coupled oscillators, denoted by $\mathcal{C}$, shown in the Appendix \ref{app:Twoosc} and the unambiguous contribution of complexity for arbitrary number of oscillators using the results of the subsection \ref{conti}, henceforth denoted by  $\mathcal{C}_1$.We will use a finite lattice for our numerical evaluation and will discretize the time steps to plot the behaviour of the circuit complexity obtained. Note that each time-dependent coefficients in these expressions explicitly depend on $\rho_k(t,\delta t)$ given by Eq.\eqref{expli}. To obtain the values of $A,B$ and $C$ which can be inserted in Eq.\eqref{expli}, we set some straightforward initial conditions. The invariant quantities, $\Omega_k$ in Eq.\eqref{gam} are taken to be $\Omega_k=1$. Using the values of $\rho_k(t,\delta t)$ for $t\rightarrow0$, we get the desired values of $A,B$ and $C$ by setting $\rho_k(0,\delta t)=1$ while $\gamma_k(0,\delta t)=0$ by using $AC-B^2=\Omega ^2$. The coupling between the oscillators is set to be $\eta=0.25$. The free parameter in the quench profile, Eq.\eqref{quenchprof} is set to $\omega_0=1$. The frequency of reference state is set to $\omega_{ref}=0.001$.

Using the exact values of circuit complexity we parameterise four different plots in this section. The first plot features the behaviour of total complexity $\mathcal{C}$ computed for two coupled oscillators discretised on a lattice of size $L=20$, varying with the quench rate $\delta t$. This plot is divided into two regions, $\delta t<1$ is the sudden quench limit (blue) region while $\delta t>1$ is the slow quench limit (yellow) region. The next three plots characterise the behaviour of unambiguous contribution of circuit complexity $C_1$ for more than two coupled oscillators discretised on a lattice of size, $L=100$, varying with dimensionless parameter $(t/\delta t)$. In these plots we mark $(t/\delta t)=1$ as Quench Point, by a dotted vertical red line. The red region for $(t/\delta t)<1.2$ features the early time behavior of circuit complexity, the yellow region $0.8<(t/\delta t)<1.2$ shows the behavior of complexity near the quench point while the blue region, $(t/\delta t)>1.2$ characterises the late time behavior of circuit complexity.\\

In Fig.\ref{fig:1}, we have plotted the numerical values of total complexity for $N=2$, coupled oscillator system with quartic interaction , using Eq.\eqref{CE2}, with respect to the quench rate $\delta t$. The blue coloured region shows the behaviour of circuit complexity in the fast regime, $\delta<<1$. In slow regime, initially, the circuit complexity monotonically increases till $\delta t=0.01$. Beyond, $\delta t=0.01$ the complexity shows a linear scaling with slope, $\log{\mathcal{C}}/\log{\delta t}=0.0825$, upto $t=0.1$. This linear scaling is then followed by a monnotonous increase in circuit complexity till $t=0.2$. Beyond $t=0.2$, circuit complexity saturates, making transition into slow regime, $\delta t>1$, marked by a yellow background. Furthermore, it is evident that, for each quartic coupling, $\lambda$, the circuit complexity shows same behaviour. However, it is clear that as one increases quartic coupling, the complexity decreases.

Although one cannot get numerical results for total circuit complexity of more than two oscillators, we have plotted the numerical values of complexity pertaining to the unambiguous block i.e., $\mathcal{C}_1$ with respect to the dimensionless parameter $(t/\delta t$) for $N=10$ coupled oscillators,  in Fig.\ref{fig:2}. The figure is divided into three regions, first is the early time behaviour for the dimensionless parameter between $0<(t/\delta t)<0.8$ marked by a red background. In this region for all the three considered quartic couplings, circuit complexity decreases linearly upto $(t/\delta t)=0.7$.  Beyond this, for $(t/\delta t)>0.7$, circuit complexity $C_1$, diverges for each particular quartic coupling $\lambda$, such that higher the value of $\lambda$, higher is the complexity $C_1$. This linear scaling is then followed by a monotonous decrease in the value of $C_1$, as we move near the quench point $(t/\delta t)=1$, marked by a yellow background. The late time behaviour, is characterised by saturation of $C_1$, for $(t/\delta t)>1.2$, which is marked by a blue background. It is evident that for different quartic couplings, $\lambda$, the circuit complexity $C_1$ scales similarly in each particular region. However as we increase the quartic coupling, $\lambda$, circuit complexity $C_1$ increases at any particular time near and beyond the quench point.

In Fig.\ref{fig:3} we have plotted the unambiguous contribution of circuit complexity $C_1$ for a discretised field theory with respect to the dimensionless parameter $(t/\delta t)$ for different number of oscillators, viz., $N=10,11,12$ (keeping quartic coupling fixed at $\lambda=0.01$). The early time behavior for a particular $N$ is characterised by a steep linear decrease in value of circuit complexity, $C_1$ upto $(t/\delta t)=0.4$. Beyond this point the complexity $C_1$ monotonically decreases and finally saturates at
$(t/\delta t)\approx0.5$. After this point, the unambiguous contribution of circuit complexity, $C_1$ remains saturated. The early time behavior of $C_1$ featuring a decrease and saturation in value of complexity is marked by red background. The behavior of $C_1$ near to the quench point as well as at late time is characterised by a constant saturated value at all times, marked by yellow and blue backgrounds respectively. It is evident that unambiguous contribution of circuit complexity, $C_1$ scales similarly for any $N$. However larger the number of oscillators $N$, larger is the value of $C_1$. In the continuum limit when we have large number of coupled oscillators, one can expect that the unambiguous contribution of circuit complexity, $C_1$ will behave similar to that shown in this figure. 

In Fig.\ref{fig:4}, we have plotted the unambiguous contribution of circuit complexity $C_1$ for a discretised field theory with respect to the dimensionless parameter $(t/\delta t)$ for different number of dimensions, viz., $d=2,3,4$ (keeping quartic coupling fixed at $\lambda=0.01$ while number of oscillators $N=10$.). For a particular $d$, the early time behavior of unambiguous contribution of circuit complexity is characterised by a monotonous decrease in the value of $C_1$, marked by a red background. The values of $C_1$ for different dimensions, $d$ begin to converge near to the quench point $(t/\delta t=1)$ and finally saturate to a constant value, this is marked by yellow background. At late times, the unambiguous contribution of circuit complexity $C_1$ remains saturated to a constant value, this is marked by a blue background. It is evident for any dimension $d$, the circuit complexity $C_1$ scales similarly. However higher the dimension $d$, larger is the value of $C_1$ at any particular time in the red region.

\section{\textcolor{Sepia}{\textbf{ \large Conclusions\label{sec5}}}}
In this article, we have studied the dynamical behavior of circuit complexity in a quenched field theory with quartic interaction. To do this, we have used a unique framework that combines Nielsen's geometric approach based on \cite{Heller,Sinha} and the invariant operator method. While preparing the reference and target states, we have employed Nielsen's geometric approach to finding the optimal circuit out of an infinite set of circuits. This approach provided a novel perspective to understand the behavior of complexity under different parametric conditions.

We have used the invariant operator method to find the exact form of eigenstates of the unperturbed part of the Hamiltonian. Combining this with perturbation theory, we were able to determine approximate solutions to the time-dependent Schrödinger equation. This enabled us to derive the analytical expression for time-dependent circuit complexity using Nielsen's method.

We have discretized the field theory on a lattice and evaluated the wavefunction for $N$-coupled oscillators having quartic perturbation and a quenched frequency. We used the ground state wavefunction to derive the analytical form of the reference and target states. By choosing a specific cost function and a minimal basis, we computed the exact form of the total complexity for two coupled oscillators and the unambiguous contribution of circuit complexity for arbitrary $N$ number of oscillators. Our study thus provides a comprehensive understanding of the dynamical behavior of circuit complexity in a quenched field theory with quartic interaction.\\

The important results of our work are appended below point-wise:
\begin{itemize} 
\item{For two coupled oscillators, we observed that, in most part of the sudden quench, the total circuit complexity monotonously increases at very small values of quench rate, then scales linearly and shows a trend of thermalisation near $\delta t=1$. In the slow quench limit, the complexity remains saturated irrespective of the quench rate. When parameterised for different values of quartic coupling, $\lambda$, it is evident that as coupling increases, even circuit complexity increases.}
\item {The exact analytical form for unambiguous contribution of the circuit complexity for $N$ coupled oscillators was derived using the results of \cite{Sinha}. The parametric variation of this circuit complexity was then plotted with respect to dimensionless parameter $(t/\delta t)$.}
\item{It is evident from these plots that unambiguous contribution of the circuit complexity decreases with respect to the dimensionless parameter. When parametrised for different quartic couplings, we find that initially the complexity decreases linearly following same line, irrespective of the quartic coupling. Near to the quench point, complexity for each quartic coupling diverges and saturates at late times. After, quench point, the complexity after divergence is clearly in direct proportion to the increasing value of quartic coupling.}
\item{We observed that, the unambiguous contribution of the circuit complexity behaves similarly, irrespective of the number of oscillators, $N$. However, as $N$ increases the respective value of complexity increases at any particular time. Using this, we commented on the conitnuum limit where the results would still be the same. }
\item{Furthermore, it is clear that the unambiguous contribution of circuit complexity for the chosen set of parameters is proportional to the increasing number of dimension at early times. However at the quench point, unambiguous contribution of circuit complexity attains a constant value irrespective of the number of dimensions and thermalises at late times.}
\end{itemize}
In conclusion, we have presented a novel approach to study the time-dependent circuit complexity in the context of quenched interacting field theory. By combining the invariant operator method and Nielsen's geometric approach, we have derived the analytical expression for circuit complexity and explored its behavior under various parametric conditions. Our results demonstrate that the quenching of the system has a significant impact on the circuit complexity of interacting field theories. Our work provides a new perspective on the understanding of the dynamical behavior of circuit complexity in interacting field theories. It also opens up avenues for future research, such as the exploration of circuit complexity in other interacting field theories and the study of its relationship with other quantum information theoretic quantities.

\textbf{Future Prospects:}
\begin{itemize}
\item {In this work we explored the effects of quantum quench on QFT with quartic coupling, in the future it would be interesting to explore the effects of quantum quench on Krylov Complexity \cite{k1,k2,k3,k4,k5} for QFT with quartic coupling. }
\item Some of the works focusing on understanding the connection between complexity and quantum entanglement are \cite{cent1,cent2,Caputa:2017ixa,cent3,cent4}. The connection between complexity and quantum entanglement in the case of quenched theories might turn out to be fruitful.
\end{itemize}

{\bf Acknowledgement:} SC would like to thank the work friendly environment of CCSP, SGT University for providing tremendous support in research and  offer the Assistant Professor (Senior Grade) position.  SC also thanks all the members of our newly formed virtual international non-profit consortium Quantum Aspects of the Space-Time \& Matter (QASTM) for elaborative discussions.  RMG, SM and NP would like to thank the members of the QASTM Forum for useful discussions. Last but not least, we would like to acknowledge our debt to the people belonging to the various parts of the world for their generous and steady support for research in natural sciences.

\onecolumngrid
\appendix
\section{\textcolor{Sepia}{\textbf{Evaluating the eigenstates of unperturbed Hamiltonian using Invariant operator method.}}}\label{app:Invar}
In this appendix, we have shown the analytical steps one can perform to evaluate the expression for the eigenstates of unperturbed Hamiltonian having a quenched (time-dependent) frequency shown in Eq.\eqref{decoup} of the section \ref{sec3}. The invariant operator method is the optimal choice to compute these eigenstates as it enables one to evaluate exact time-dependent wavefunctions by solving the time-dependent Schrodinger equation. A detailed analysis on the invariant operator method can be referred from \cite{1994NuPhB.424..443H}.\\
First, we define the creation $(a^\dagger_k)$ and annihilation $(a_k)$ operators given by,
\begin{equation}\label{crea}
\begin{aligned}
 a_k=\frac{1}{\sqrt{2\dot{\gamma_k}}}\biggr[\dot{\gamma}_k\biggr(1-i\frac{\dot{\rho_k}}{\rho_k\dot{\gamma_k}}\biggr)\tilde{x}_k+i\tilde{p}_k\biggr] \\
 a_k^\dagger=\frac{1}{\sqrt{2\dot{\gamma_k}}}\biggr[\dot{\gamma}_k\biggr(1+i\frac{\dot{\rho_k}}{\rho_k\dot{\gamma_k}}\biggr)\tilde{x}_k-i\tilde{p}_k\biggr].\\
\end{aligned}
\end{equation}
Here, $k=0,1,\cdots,N-1$. Further, $\gamma_k$ and $\rho_k$ are time-dependent factors while $\dot{\gamma}_k=\partial_T\gamma_k$, $\dot{\rho}_k=\partial_T\rho_k$ and $\ddot{\rho}_k=\partial_T^2\rho_k$.  One can show that the operators in Eq.\eqref{crea} satisfy the commutation relation $[a_i,a_j^\dagger]=\delta_j^i$. We fix the time-dependent factor $\rho_k$ as the solution to the Ermakov-Milne-Pinney equation for each oscillator,
\be\label{EMP1}
\Ddot{\rho_k}+\omega_k^2\rho_k=0,
\ee
where $\omega_k$ denotes frequency for each coupled oscillator Eq.\eqref{frqe}. If we define,
\be\label{EMP2}
\alpha_k=\omega_0^2+\frac{4\eta^2\left(\sin^2\left(\frac{\pi  i}{N}\right)\right)}{\tanh^2\left(T\right)},
\ee
for,  $k=0,1,\cdots,N-1$, then using Eq.\eqref{quenchprof} one can rewrite Eq.\eqref{EMP1} as:
\be\label{EMPf}
\Ddot{\rho_k}+\alpha_k\tanh^2(T)\rho_k=0.
\ee
Here $k=0,1,\cdots,N-1$. We assume that the solution to Eq.\eqref{EMPf} is of the form:
\be\label{EMPs}
\rho_k(t,\delta t)=c_1\epsilon^1_k(t,\delta t)+c_2\epsilon^2_k(t, \delta t).
\ee
Here $c_1$ and $c_2$ represent numerical constants. On the other hand $\epsilon^1_k$ and $\epsilon^2_k$ are two complex valued coefficients for each $k=0,\cdots,N-1$. However we can consider only the term with, $\epsilon^1_k$ as one of the solutions to Eq.\eqref{EMPf}, by setting $c_2=0$. Using Mathematica one can compute the exact form of $\epsilon^1_k$, following the steps shown in \cite{QE1}, which is mentioned below:
\be\label{EMPs1}
\epsilon^1_k=\left(e^{\frac{2t}{\delta t}}\right)^{-\frac{1}{2} i \delta t \alpha_k} \left(e^{\frac{2t}{\delta t}}+1\right)^{\frac{1}{2} \left(\sqrt{1-4 \delta t^2 \alpha_k^2}+1\right)} \, _2F_1\left(\frac{1}{2} \left(\sqrt{1-4 \delta t^2 \alpha_k^2}+1\right),\frac{1}{2}\left(\sqrt{1-4 \delta t^2 \alpha_k^2}-2 i\delta t \alpha_k^2+1\right);1-i\delta t\alpha_k^2;-e^{\frac{2 t}{\delta t}}\right),
\ee    
where, $_2F_1$ denotes the hyper-geometric function.
These complex valued solutions can be rewritten in the form $\epsilon^1_k=\varepsilon_k+i\zeta_k$, where $\varepsilon_k$ and $\zeta_k$ can be treated as real and linearly independent equations; for each $k=0,\cdots,N-1$. As evaluated in \cite{2016arXiv160308747M}, the solution to Eq.\eqref{EMPf} is guaranteed to be of the form:
\be\label{expli}
\rho_k(t,\delta t)=\sqrt{A\varepsilon_k^2(t,\delta t)t+2B\varepsilon(t,\delta t)\zeta_k(t,\delta t)+C\zeta_k^2(t,\delta t)}~.
\ee
Also, $\Omega_k=\rho_k^2\dot{\gamma_k}$, is an invariant quantity with respect to time. Hence, $\gamma_k$ can be computed as follows,
\be\label{gam}
\gamma_k(t,\delta t)=\int^t_0\frac{\Omega_k}{\rho_i^2(t,\delta t)}dt~~.
\ee
Note that all the quantities are now a function of $t$ and $\delta t$, this time-dependence is however suppressed along the most part of this article; until necessary. The creation and annihilation operators in Eq.\eqref{crea} can now be used to construct invariant operator, for each of the N decoupled Hamiltonians, of Eq.\eqref{decoup}:
\bea\label{invar}
I_k=\Omega_k\biggr(a_k^\dagger a_k+\frac{1}{2}\biggr).
\eea
Here, $k=0,1\cdots,N-1$. Assuming that each invariant operator $I_k$ is one of the complete set of commuting observables for the respective Hamiltonian $H_k$ assures that each $I_k$ has its own eigenstates. The ground state for this spectrum of each invariant operator can be computed using annihilation operator of Eq.\eqref{crea} by solving, $a_ku_{0_k}=0$. The expression for the ground state of for the spectrum of each invariant operator $I_k$ of Eq.\eqref{invar}, for each $k=0,1\cdots,N-1$ is given by,
\bea
u_{0_k}=\biggr(\frac{\dot{\gamma}_k}{\pi}\biggr)^{1/4}\exp\biggr[-\frac{\dot{\gamma_k}}{2}(1-i\dot{\rho_k}\rho_k\dot{\gamma_k})\tilde{x}_k^2\biggr].
\eea
The creation operator in Eq.\eqref{crea} can then be used to evaluate the $n^{th}$ eigenstate of the invariant-operator $I_k$ which is given by,
 \bea\label{eigenI}
u_{n_k}=\frac{1}{\sqrt{n!}}(a_k^\dagger)^nu_{0_k}=\biggr(\frac{1}{2^{n_k}n_k!}\biggr)\biggr(\frac{\dot{\gamma_k}}{\pi}\biggr)^{1/4}
\exp\biggr[\dot{\gamma}_k\biggr(1-\frac{i\dot{\rho_k}}{\rho_k\dot{\gamma_k}}\biggr)\tilde{x}k^2\biggr]
\textbf{H}_{n_k}\left[\sqrt{\dot{\gamma_k}}\tilde{x}_k\right].
\eea   

Here $\textbf{H}_{n_k}$ denotes the Hermite polynomial of order $n_k$; for each $k=0,\cdots,N-1$.
The eigenstates of the invariant operator shown in Eq.\eqref{eigenI} can now be used to compute the wavefunction for each decoupled Hamiltonian of Eq.\eqref{decoup} such that, $\psi_{n_k}=e^{i\beta_{n_k}}u_{n_k}$, where $\beta_{n_k}=-({1}/{2}+n_k)$; for $k=0,\cdots,N-1$.

\section{\textcolor{Sepia}{\textbf{ \large$\mathcal{C}_{\kappa=1}^{(2)}$ in terms of renormalized parameters}}}\label{app:Renorm}

As discussed in \cite{Sinha} one can attempt to find the form of $\mathcal{C}_{\kappa=1}^{(2)}$ by using renormalisation. In this subsection we give an outline of the modified expression for the same in case of a quenched interacting field theory discretised on a $d-$ dimensional lattice containing $N$ oscillators.\\
The renormalized matrix elements for $A_2$ block are given as \cite{Sinha},
\be
A_2[m,n]=\frac{a_{mn}\lambda_R\delta^{-d}}{V^{\frac{1}{d-1}}f(\tilde{\omega}_i)}.
\ee
Note that all these elements are time-dependent due to the form of frequency, $\tilde{\omega}_i$ chosen as a quench profile. The eigenvalues then take the general form \cite{Sinha},
\be\label{CE1}
\Lambda_i^{(2)}=\frac{b_j\lambda_R\delta^{-d}}{V^{\frac{1}{d-1}}g(\tilde{\omega}_i)},
\ee
where $j\in \{0,1,\cdots (\textit{Dim} A_2-1\}$ while $i\in \{0,1,\cdots N-1\}$. As shown in \cite{Sinha} the renormalized penalty factor is,
\be
\mathcal{A}=(\lambda_R\delta^{4-d})^\mu\delta^{-\upsilon}V^{\frac{\upsilon}{d-1}}.
\ee
Here $\upsilon$ and $\mu$ are arbitrary integers which can be fixed by using physical arguments as discussed in \cite{Sinha}. Using Eq.\eqref{CE}, the renormalized-complexity contribution for $A_2$ block in $d-$ dimensions can then be written as,
\be
\mathcal{C}_{\kappa=1}^{(2)}=\frac{(\lambda_R\delta^{4-d})^\mu\delta^{-\upsilon}V^{\frac{\upsilon}{d-1}}}{2}\sum_{k=0}^{d-1}\sum_{i_k=0}^{(\text{Dim}A_2)-1}\Big|\log\Big(\frac{\Lambda_{i_k}^{(2)}\delta^2}{h_i\tilde{\omega}_{ref}\lambda_0}\Big) \Big|.
\ee
Note that now each $i_k\in\{1,2,\cdots(\text{Dim}A_2-1)\}$ for $k\in\{1,2,\cdots d-1\}$. Using the renormalized form of eigenvalues from Eq.\eqref{CE1} we finally obtain,
\be
\mathcal{C}_{\kappa=1}^{(2)}=\frac{(\lambda_R\delta^{4-d})^\mu\delta^{-\upsilon}V^{\frac{\upsilon}{d-1}}}{2}\sum_{k=0}^{d-1}\sum_{i_k=0}^{(\text{Dim}A_2)-1}\Big|\log \Big(\frac{b_{i_k}\lambda_R\delta^{2-d}}{V^{\frac{1}{d-1}}g(\tilde{\omega}_i)h_{i_k}\tilde{\omega}_{ref}\lambda_0}\Big)\Big|
\ee.
Note that, we have not used this expression for getting numerically exact results.

\section{\textcolor{Sepia}{\textbf{ \large Circuit Complexity for two oscillators\label{app:Twoosc}}}}
In the subsection \ref{sub:R/T}, we commented on the ambiguities in fixing the coefficients in the ambiguous $A_2$ block for arbitrary number of oscillators, $N$. Due to these ambiguities one cannot obtain numerical results for contribution from $A_2$ block in circuit complexity for $N$ oscillators. However, in this appendix,  we choose a minimal basis for the case of two oscillators and hence get rid of these ambiguities to get the total contribution of $A_1$ as well as $A_2$ block in circuit complexity. We begin by specialising the wavefunction (in normal modes) in Eq.\eqref{T1} to that of two oscillators by inserting $N=2$, written as:
\begin{align}\label{w2}
  \psi_{0,0}\left(\bar{x}_{0}, \bar{x}_{1}\right)=\frac{\left(g_{0} g_{1}\right)^{1/ 4}}{\sqrt{\pi}}\exp\left[-\iota(\gamma_0+\gamma_1)\right]\exp\left[C_{0}\right] \exp \Big[&-\frac{1}{2}\Big(C_{1}  \bar{x}_{0}^{2}+C_{2}  \bar{x}_{1}^{2}+C_{3}  \bar{x}_{0}^{2}  \bar{x}_{1}^{2}+C_{4}  \bar{x}_{0}^{4}+C_{5}  \bar{x}_{1}^{4}\Big)\Big].
\end{align}
The exact form of $C_i$ for $i=0$ to $i=5$ can be inferred from the table in the Appendix \ref{app:Table}. Next, we write the above wavefunction in the following form:
\begin{equation}
\psi^{s}(\bar{x}_{0}, \bar{x}_{1})=\mathcal{N}^{s} \exp \Big[-\frac{1}{2}\Big(v_{a}  A(s)_{a b} \ v_{b}\Big)\Big].
\label{Eqdd}
\end{equation}
Here $\mathcal{N}^{s}$ is the normalisation factor.
Similar to that of generalised wavefunction inserting $s=1$ in the above equation will correspond to the target state, while inserting $s=0$ corresponds to the reference state. We choose an unentangled and non-Gaussian reference state given by,
\begin{align}\nonumber
\psi^{s=0}(\bar{x}_{0}, \bar{x}_{1})= &\mathcal{N}^{s=0}\exp \Big[-\frac{\tilde{\omega}_{r e f}}{2}(\bar{x}_{0}^{2}+\bar{x}_{1}^{2}+\frac{\lambda_{0}}{2}(\bar{x}_{0}^{4}+\bar{x}_{1}^{4}+6 \bar{x}_{0}^{2} \bar{x}_{1}^{2}))\Big].
\label{ref2}
\end{align}
Here $\lambda_{0}$ parameterizes the non-Gaussianity of the reference state. The exponential in the above equation can be written in form of a matrix $A(s=0)$ by choosing a basis,
\begin{equation}\label{bais}
  \vec{v}=\{\bar{x}_{0}, \bar{x}_{1}, \bar{x}_{0} \bar{x}_{1}, \bar{x}_{0}^{2}, \bar{x}_{1}^{2}\}.
\end{equation}
The matrix then takes the following form:
\begin{equation}\label{Eq_3.14}
A(s=0)=\left(\begin{array}{ccccc}
A_{1}^0 & 0   \\
0 & A_{2}^0  \\
\end{array}\right),
\end{equation}
where,
\begin{equation}
A_{1}^0= \left( \begin{array}{cc}
\tilde{\omega}_{ref} & 0 \\
0 & \tilde{\omega}_{ref}
\end{array} \right)~~~~~~~~~~;
\hspace{2cm}
A_{2}^0=\lambda_{0} \tilde{\omega}_{ref}\left( \begin{array}{ccc}
b & 0 &0\\
0 & \frac{1}{2}&\frac{1}{2}(3-b)\\
0&\frac{1}{2}(3-b)&\frac{1}{2}
\end{array} \right). \nonumber
\end{equation}
One can choose the values of $b$ such that matrix $A_{2}^0$ is non-singular. To diagonalize $A_{2}^0$ we set $b=3$. On the other hand to choose a non-Guassian reference state we set $\lambda_0=1.5$.\\
Similarly one can get the target state in form of matrix $A(s=1)$ given by,
\begin{equation}
\psi^{s}(\bar{x}_{0}, \bar{x}_{1})=\mathcal{N}^{s=1} \exp \Big[-\frac{1}{2}\Big(v_{a}  A(s=1)_{a b} \ v_{b}\Big)\Big].
\label{Eq_3.10}
\end{equation}
Choosing the same basis as that in Eq.\eqref{bais} we can write the matrix for target state as,
\begin{equation}\label{Eqdd1}
A(s=1)=\left(\begin{array}{ccccc}
A_{1}^1 & 0   \\
0 & A_{2}^1  \\
\end{array}\right),
\end{equation}
where,
\begin{equation}
A_{1}^1= \left( \begin{array}{cc}
C_1 & 0 \\
0 & C_2
\end{array} \right)~~~;
\hspace{2cm}
A_{2}^1=\left( \begin{array}{ccc}
\tilde{b}C_5 & 0 &0\\
0 & C_3&\frac{1}{2}(1-\tilde{b})C_5\\
0&\frac{1}{2}(1-\tilde{b})C_5&C_4
\end{array} \right). \nonumber
\end{equation}
The parameter $\tilde{b}$ can be chosen such that $A_2^1$ is non-singular. Further to diagonalise $A_2^1$ we set $\tilde{b}=1$. It is clear that with a minimal choice of basis given in Eq.\eqref{bais} one can fix all the elements of both $A_1$ and $A_2$ blocks for the case of two oscillators. Aimed with the expression for circuit complexity for $N$ oscillators given in Eq.\eqref{CE}, and setting the penatly factor $\mathcal{A=1}$, we have computed the circuit complexity for the two quenched oscillators with quartic coupling, by using:  
\be\label{CE2}
 \mathcal{C}_{\kappa=1}=\frac{1}{2}\Big(\log\Big|\frac{\text{det}\hspace{0.05cm}A^1_1}{\text{det}\hspace{0.05cm}A^0_1}\Big|+\log\Big|\frac{\text{det}\hspace{0.05cm}A^1_2}{\text{det}\hspace{0.05cm}A^0_2}\Big|\Big).
 \ee
Here $det$ refers to the determinant of the respective block of the matrix. Note that the quench profile chosen as the frequency scale of the system imposes time-dependence on each element of the target state matrix block viz., $A_1^1$ and $A_2^1$. The complexity therefore becomes time-dependent, this is clearly shown in the numerical results discussed in the section \ref{sec4}.

\section{Tabulated Values of Coefficients}\label{app:Table}
In this appendix, the values of various coefficients we have used in some steps to compute the analytical expression of circuit complexity, are tabulated in respective tables.\\
\begin{itemize}
\item We begin by listing the values of $B_i$ for $i=1,2...5$ in equation Eq.\eqref{PC1} of section \ref{perturbed} in the table given below.

\begin{longtable}{|c|c|}
\hline
\rowcolor{Gray}
 $B_{i}$ & Coefficient of $B_{i}$\\
\hline
$B_{1}(a)$ &
{\Large \scalebox{0.70}{
\parbox[t]{20cm}
{\Large $\\-\frac{3 x_a^2}{4 N g_a W_a}+\frac{9}{16 N g_a^2 W_a}-\frac{x_a^4}{4 N W_a}
\\$}
}}\\ \hline
$B_{2}(b,c)$ &
\scalebox{0.70}{
\parbox[t]{20cm}
 {\Large $\\-\frac{3 x_b^2 W_c}{2 N W_b g_c \left(W_b+W_c\right)}-\frac{3 W_b x_c^2}{2 N g_b W_c \left(W_b+W_c\right)}+\frac{3}{4 N g_b g_c \left(W_b+W_c\right)}+\frac{3 W_c}{4 N g_b W_b g_c \left(W_b+W_c\right)}+\frac{3 W_b}{4 N g_b g_c W_c \left(W_b+W_c\right)}-\frac{3 x_b^2 x_c^2}{N \left(W_b+W_c\right)}
 \\$}
 }\\ \hline
$B_{3}(d,e)$ &
\scalebox{0.70}{
\parbox[t]{20cm}
{\Large $\\-\frac{12 x_d W_e x_e}{N g_e \left(W_d+W_e\right) \left(W_d+3 W_e\right)}-\frac{4 x_d x_e^3}{N \left(W_d+3 W_e\right)}
\\$}
 }\\ \hline

$B_{4}(f,m,h)$ &
\scalebox{0.70}{
\parbox[t]{20cm}
 {\Large  $\\-\frac{12 x_f x_h W_m}{N g_m \left(W_f+W_h\right) \left(W_f+W_h+2 W_m\right)}-\frac{12 x_f x_h x_m^2}{N \left(W_f+W_h+2 W_m\right)}
\\$}
 }\\ \hline

$B_{5}(i,j,k,l)$ &
\scalebox{0.70}{
\parbox[t]{20cm}
 {\Large $\\-\frac{24 x_i x_j x_k x_l}{N \left(W_i+W_j+W_k+W_l\right)}
\\
$}
 }\\ \hline\hline\hline
\end{longtable}
Note that $g_k$ for $k=0,\cdots N-1$ is defined in the subsection \ref{subsec3.1}, while $W_k$ in Eq.\eqref{eigent}. Also all the indices in the tabulated expressions run from $0$ to $N-1$.

\item Next, we tabulate the values of coefficients $C_i$ for $i=1,2...5$ in equation Eq.\eqref{w2} of the Apendix \ref{app:Twoosc}.

\begin{longtable}{|c|c|}

\hline
\rowcolor{Gray}
 $C_{i}$ & Coefficient of $C_{i}$\\
\hline
$C_{0}$ &

{\Large \scalebox{.70}{
\parbox[t]{20cm}
{ $\\\frac{9 \lambda }{32 g_0^2 W_0}+\frac{9 \lambda }{32 g_1^2 W_1}+\frac{3 \lambda }{8 g_0 g_1 \left(W_0+W_1\right)}+\frac{3 \lambda  W_1}{8 g_0 g_1 W_0 \left(W_0+W_1\right)}+\frac{3 \lambda  W_0}{8 g_0 g_1 W_1 \left(W_0+W_1\right)}\\$}}
 }\\ \hline
$C_{1}$ &

{\Large \scalebox{0.70}{
\parbox[t]{20cm}
 {$\\-\frac{3 \lambda }{8 g_0 W_0}-\frac{3 \lambda  W_1}{4 g_1 W_0 \left(W_0+W_1\right)}-\frac{\nu _0}{2}
 \\$}}
 }\\ \hline
$C_{2}$ &
\scalebox{0.70}{
\parbox[t]{20cm}
 {\Large $\\-\frac{3 \lambda }{8 g_1 W_1}-\frac{3 \lambda  W_0}{4 g_0 W_1 \left(W_0+W_1\right)}-\frac{\nu _1}{2}
 \\$}
 }\\ \hline
$C_{3}$ &
\scalebox{0.70}{
\parbox[t]{20cm}
{\Large $\\-\frac{\lambda }{8 W_0}
\\$}
 }\\ \hline
$C_{4}$ &
\scalebox{0.70}{
\parbox[t]{20cm}
 {\Large  $\\-\frac{\lambda }{8 W_1}
\\$}
 }\\ \hline
$C_{5}$ &
\scalebox{0.70}{
\parbox[t]{20cm}
 {\Large $\\-\frac{3 \lambda }{2 \left(W_0+W_1\right)}
 \\
$}
 }\\ \hline
\hline\hline \hline
\end{longtable}
Note that $g_k=\dot{\gamma}_k$ where $\gamma_k$ for $k=0,1$ can be computed by solving the EMP equation as shown in the section \ref{sec1}, while $W_k$ can be computed using Eq.\eqref{eigent}.

\twocolumngrid

\end{itemize}

\bibliography{referencesnew}
\bibliographystyle{utphys}

\end{document}